\documentclass[twocolumn,pra,showpacs]{revtex4}

\usepackage{amsmath}
\usepackage{graphicx}
\usepackage{amsfonts}
\usepackage{amssymb}

\begin{document}

\title{Atomic quantum memory: cavity \textit{vs} single pass schemes}
\author{A. Dantan\footnote[3]{email: dantan@spectro.jussieu.fr}}

\author{A. Bramati}

\author{M. Pinard}

\affiliation{Laboratoire Kastler Brossel, Universit\'{e} Pierre et
Marie Curie,\\
Case 74, 4 place Jussieu, 75252 Paris Cedex 05, France}

\date{\today}

\newcommand{\beq}{\begin{equation}}
\newcommand{\eeq}{\end{equation}}
\newcommand{\beqr}{\begin{eqnarray}}
\newcommand{\eeqr}{\end{eqnarray}}
\newcommand{\lb}[1]{\label{#1}}
\newcommand{\ct}[1]{\cite{#1}}
\newcommand{\dt}{\frac{\partial}{\partial t}}
\newcommand{\dz}{\frac{\partial}{\partial z}}
\newcommand{\nn}{\nonumber}

\begin{abstract}
This paper presents a quantum mechanical treatment for both atomic
and field fluctuations of an atomic ensemble interacting with
propagating fields, either in Electromagnetically Induced
Transparency or in a Raman situation. The atomic spin noise spectra
and the outgoing field spectra are calculated in both situations.
For suitable parameters both EIT and Raman schemes efficiently
preserve the quantum state of the incident probe field in the
transfer process with the atoms, although a single pass scheme is
shown to be intrinsically less efficient than a cavity scheme.
\end{abstract}

\maketitle

\section{\bigskip Introduction}

There has recently been a lot of interest in quantum communication
at the light-atom interface, with the prospect of realizing quantum
information networks composed of nodes of atomic ensembles connected
by light \ct{lukin,duan}. A basic requirement of such networks is
the ability to perform quantum state exchanges between fields and
atoms. There have been various proposals to write, store and readout
a field state onto a long-lived atomic spin, and several experiments
have already demonstrated the possibility to manipulate quantum
states between field and atoms: on the one hand "slow-light"
experiments based on Electromagnetically Induced Transparency (EIT)
\ct{harris} have shown that a pulse of light could be stored and
retrieved inside an atomic cloud \ct{hau}. In the weak probe regime
the conservation of the quantum character of the pulse was predicted
using the concept of "dark-state polaritons" \ct{fleischhauer}, but
remains to be demonstrated experimentally. On the other hand
Raman-type interactions have been used to entangle two atomic
ensembles and map a polarization state of light onto an atomic spin
\ct{julsgaard}, and the mapping and storage of coherent states have
been reported very recently by Julsgaard \textit{et al.}
\ct{julsgaard2}. There are also several proposals to realize a
quantum memory using such a scheme \ct{kuzmich}. In recent works we
studied how non-classical light states could be transferred to
atoms, and predicted quasi-ideal quantum state transfer between
field and atoms placed in an optical cavity, in both EIT and Raman
configurations \ct{dantan1,dantan2}. In particular we showed that
squeezed states and Einstein-Podolsky-Rosen states (EPR) could be
mapped onto atomic ground state spins with a high efficiency. We
also developed a method to readout the atomic state in the field
exiting the cavity, thus allowing quantum memory operations in a
controlled and efficient manner. In these calculations the cavity
plays an important part to improve the collective atom-field
coupling which scales linearly with the number of atoms. Moreover,
the intrinsic noise coming from spontaneous emission or ground state
decoherence is substantially damped by the cavity interaction,
allowing in principle quasi-reversible quantum state exchanges
between the field and the atoms. In the present paper we extend
these cavity results to single pass interaction, and show that good
quality quantum state transfers are also possible either in EIT and
Raman situations. We first present a general method to calculate
both the field and the atomic noise spectra in a one-dimensional
propagation problem. We then apply it to the case of squeezed vacuum
input field state, derive the outgoing field spectrum and the atomic
variances, first in EIT and then in Raman. We analyze the mapping
efficiencies, the effect of ground state decoherence, and compare
the results obtained in the single pass schemes with those of the
cavity schemes. An important result is that, in any situation, the
efficiency increases faster with the number of atoms in a cavity
scheme ($\varpropto N$) than in a single pass scheme ($\varpropto
\sqrt{N}$).

\section{Single pass scheme}

In the following sections we address the issue of one-dimensional
field propagation through a dilute atomic cloud of length $L$,
cross-section $\mathcal{S}$ and containing $N$ atoms uniformly
distributed. We assume that the atomic cloud is elongated with
Fresnel number of order unity, so that the emission can be
considered one-dimensional \ct{raymer}. In order to take into
account transverse effects a three-dimensional theory would be
required \ct{duan2}, which is beyond the scope of the present paper.
In the first section we introduce continuous operators by dividing
the atomic medium into transverse slices, as in \ct{fleischhauer95},
and we give the atom-field evolution equations. In the next sections
we study the continuous interaction of a coherent pump field and a
squeezed vacuum probe field with the atoms, and calculate the
spectra of the field exiting the cloud in two situations: in EIT -
both fields are one- and two-photon resonant - and in a Raman
configuration - large one-photon detunings, but two-photon resonance
is maintained. These situations have been shown to be the most
favorable to the conservation of quantum states during atom-field
transfer operations \ct{dantan1,dantan2,dantan3}. For each
configuration we also calculate the atomic ground state coherence
variances and show that the incident field state can be perfectly
mapped onto the atoms.

\subsection{Atom-fields evolution equations}

In order to treat the paraxial propagation problem we write the
positive frequency component of the copropagating electric fields
$E_j$ ($j=1,2$) as $E^{(+)}_j(z,t)=\mathcal{E}_{0j}
A_j(z,t)e^{i(kz-\omega_jt)}$, where $\omega_j$ is the laser
frequency,
$\mathcal{E}_{0j}=\sqrt{\hbar\omega_j/2\epsilon_0\mathcal{S}L} $ and
$A_j(z,t)$ is a dimensionless slowly-varying envelope operator,
satisfying
\beqr\nn[A_j(z,t),A_j^{\dagger}(z',t')]=\frac{L}{c}\delta\left(t-t'-(z-z')/c\right).\eeqr
From the single-atom operators $\sigma_{\mu\nu}^{j}(t)$ (in the
rotating frame of their laser frequency) one can define continuous
operators at position $z$ by averaging on a slice of length $\Delta
z$ \ct{fleischhauer95} \beqr\nn \sigma_{\mu\nu}(z,t) =
\frac{L}{N\Delta z}\sum_{z\leq z^j\leq z+\Delta z}
\sigma_{\mu\nu}^j(t).\eeqr Denoting the control field by $A_1$ and
the probe field by $A_2$ the interaction Hamiltonian can then be
expressed as \beqr\nn H=-\hbar\sum_{j=1,2}
\int\frac{dz}{L}N[g_jA_j(z,t)\sigma_{3j}(z,t)+h.c.],\eeqr with
$g_j=d_j\mathcal{E}_{0j}/\hbar$ the atom-field coupling constants.
The field evolution equations are obtained from Maxwell's
propagation equations in the slowly-varying envelope approximation
\beqr\left(\dt+c\dz\right)A_j(z,t)=ig_jN\sigma_{j3}(z,t)\hspace{0.5cm}(j=1,2)\eeqr
The evolution equations for the atomic variables are given by a set
of Heisenberg-Langevin equations \beqr
\nn\dt\sigma_{13}&=&-(\gamma+i\Delta_1)\sigma_{13}+ig_1A_1(\sigma_{11}-\sigma_{33})+ig_2A_2\sigma_{21}+f_{13}\\
\nn\dt\sigma_{23}&=&-(\gamma+i\Delta_2)\sigma_{23}+ig_2A_2(\sigma_{22}-\sigma_{33})+ig_1A_1\sigma_{12}+f_{23}\\
\nn\dt\sigma_{21}&=&-(\gamma_0-i\delta)\sigma_{21}+ig_1A_1^{\dagger}\sigma_{23}-ig_2A_2\sigma_{31}+f_{21}\\
\nn\dt\sigma_{11}&=&-\gamma_0\sigma_{11}+\gamma\sigma_{33}+\Lambda_1+ig_1A_1^{\dagger}\sigma_{13}-ig_1A_1\sigma_{31}+f_{11}\\
\nn\dt\sigma_{22}&=&-\gamma_0\sigma_{22}+\gamma\sigma_{33}+\Lambda_2+ig_2A_2^{\dagger}\sigma_{23}-ig_2A_2\sigma_{32}+f_{22}\\
\nn\dt\sigma_{33}&=&-2\gamma\sigma_{33}-(ig_1A_1^{\dagger}\sigma_{13}-ig_1A_1\sigma_{31})\\
\nn&&-(ig_2A_2^{\dagger}\sigma_{23}-ig_2A_2\sigma_{32})+f_{33}\eeqr
where the $\Delta_i$'s are the detunings from resonance,
$\delta=\Delta_1-\Delta_2$ is the two-photon detuning, $\gamma$ the
optical dipole decay rate (taken equal on both transitions for
simplicity) and $\gamma_0$ the decay of the ground state coherence,
modeling collisions or accounting for the transit of the atoms
outside the interaction area with the light beams. The $\Lambda_i$'s
are chosen to maintain the total number of atoms constantly equal to
$N$. The $f_{\mu\nu}$'s are $\delta$-correlated Langevin operators,
the correlation functions of which are of the form \beqr\nn \langle
f_{\mu\nu}(z,t)f_{\rho\sigma}(z',t')\rangle=\frac{L}{N}D_{\mu\nu\rho\sigma}\delta(t-t')\delta(z-z')\eeqr
The diffusion coefficients $D_{\mu\nu\rho\sigma}$ can be calculated
via the quantum regression theorem.

\subsection{EIT interaction}

In the so-called EIT situation ($\Delta_1=\Delta_2=\delta=0$), the
presence of the control field allows the probe field to propagate
with little dissipation. Correlations between pump and probe fields
have been investigated \ct{fleischhauer95} and observed
\ct{nusszenveig}, and, recently, the quantum character of a squeezed
vacuum probe has been shown to be partially conserved in EIT
\ct{akamatsu}, but little attention has been paid to the atomic
variables. We will focus on the case of a coherent pump field and a
zero-mean valued probe field with some quantum fluctuations over a
broad bandwidth, e.g. squeezed vacuum. In such a situation all the
atoms are pumped into level 2 in steady state and the fluctuations
of both fields are decoupled \ct{dantan1}. Moreover, the
fluctuations of the probe field are only coupled to the atomic
ground state coherence and the optical coherence $\sigma_{23}$.
Linearizing around this steady state one obtains for the
fluctuations of the probe field amplitude quadrature
$X=A_2+A_2^{\dagger}$ \beqr\label{Xzt}
\left(\dt +c\dz\right)X(z,t) &=& -2gN\sigma_y(z,t)\\
\label{sigmayzt}\dt \sigma_y(z,t) &=& -\gamma\sigma_y+\Omega
j_x+\frac{g}{2} X+f_{\sigma_y}\\\label{jxzt} \dt j_x(z,t) &=&
-\gamma_0 j_x-\Omega \sigma_y+f_{j_x},\eeqr where $\Omega=g_1\langle
A_1\rangle$ is assumed real, $\sigma_y=(\sigma_{23}-\sigma_{32})/2i$
and $j_x=(\sigma_{21}+\sigma_{12})/2$ are the fluctuations of the
optical dipole and ground state coherence. Similar equations relate
the field
phase quadrature, the other atomic components $\sigma_x$ and $j_y$.\\

Fourier transforming these equations, one derives the outgoing field
spectrum $S_{X^{out}}(\omega)$, defined by \beqr \langle
X^{out}(\omega)
X^{out}(\omega')\rangle=2\pi\delta(\omega+\omega')\frac{L}{c}S_{X^{out}}(\omega).\eeqr
Assuming that the incident amplitude squeezing spectrum is constant
and equal to $S_{X^{in}}$ over the frequency bandwidth considered,
one gets \beqr\label{speit}
S_{X^{out}}(\omega)=1-\left[1-S_{X^{in}}\right]e^{-(\alpha+\alpha^*)},\eeqr
with \beqr\nn \alpha(\omega)=-i\frac{\omega
L}{c}+C\frac{\gamma(\gamma_0-i\omega)}{(\gamma-i\omega)(\gamma_0-i\omega)+\Omega^2}.\eeqr
We have denoted by $\Gamma_E=\Omega^2/\gamma$ the optical pumping at
resonance and introduced a cooperativity parameter
\ct{dantan1}\beqr\nn C=\frac{g^2N}{\gamma}\frac{L}{c}.\eeqr The term
in $i\omega L/c$ corresponds to the field dephasing due to the
propagation in vacuum. However, in EIT conditions, the propagation
is strongly modified: expanding $\alpha(\omega)$ around
zero-frequency yields the well-known result \beqr\nn
\alpha(\omega)=A-i\omega\frac{ L}{v_g}+\mathcal{O}(\omega^2)\eeqr
where \beqr\label{A}
A&=&C\frac{\gamma\gamma_0}{\gamma\gamma_0+\Omega^2},\\\label{vg}
v_g&=&\frac{c}{1+g^2N\frac{\Omega^2-\gamma\gamma_0}{(\Omega^2+\gamma\gamma_0)^2}},\eeqr
represent the absorption of the field at zero-frequency and the
group velocity change in EIT, which is drastically reduced when
$\gamma\gamma_0\ll\Omega^2\ll g^2N$ \ct{fleischhauer}.
\begin{figure}[h]
\includegraphics[width=8cm]{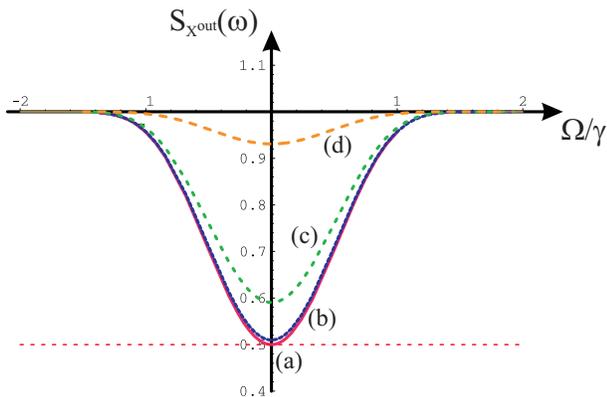}
\caption{Outgoing field squeezing spectrum in EIT for different
values of $\gamma_0$: (a) 0, (b) $\gamma/1000$, (c) $\gamma/100$,
(d) $\gamma/10$. Parameters: $\Gamma_E=10\gamma$, $C=100$.}
\label{fig1}
\end{figure}
A typical spectrum is plotted in Fig. \ref{fig1} for an initial
amplitude squeezing of 3 dB and different values of $\gamma_0$: the
interesting result is that the outgoing field is squeezed only in a
certain transparency window, of width \beqr\nn\Delta\omega\simeq
\Gamma_E\sqrt{\frac{\ln
2}{2C}\left(1-\frac{C\gamma_0}{\Gamma_E}\right)}\eeqr when $C\gg 1$
and $\Gamma_E\gg\gamma_0$. Outside this window the outgoing field
fluctuations are "absorbed" by the atoms and the field is at the
shot noise. Besides, the more the atoms, the larger $C$ and the
narrower the transparency bandwidth is. Last, an important parameter
is $\gamma_0$, which, although it can be made very small with
respect to the optical pumping and the spontaneous emission rates,
can be responsible for a substantial squeezing reduction at low
frequency when the number of
atoms is large [see Fig. \ref{fig1} and Eq. (\ref{A})].\\

It is also very interesting to look at what happens to the atoms. As
conjectured by Fleischhauer and Lukin \ct{fleischhauer}, and
predicted in a cavity configuration in \ct{dantan1}, the atomic
coherence may be squeezed by almost the same amount as the incident
field for a good choice of the interaction parameters. The atoms are
said to be spin-squeezed when the variance of one spin component in
the plane orthogonal to the mean spin is less than its coherent
state value. More precisely, we define collective atomic observables
by integrating the continuous operators over the cloud length
\beqr\nn J_{\mu}(t)=N\int\frac{dz}{L}j_{\mu}(z,t).\eeqr In our case
the collective mean spin is completely polarized along $z$: $\langle
J_z\rangle=N/2$, and, for a coherent spin state, one has $\Delta
J^2_x=\Delta J^2_y=N/4$. A spin-squeezed ensemble will have $\Delta
J^2_{\theta}<N/4$ for some component $J_{\theta}$ in the
$(x,y)$-plane \ct{ueda}.

\begin{figure}[h]
\includegraphics[width=8cm]{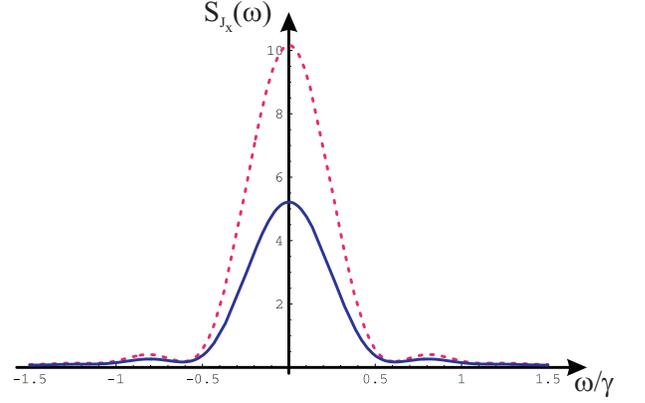}
\caption{Noise spectrum of $x$-component of the spin in EIT when the
incident field is in a coherent state [$S_{X^{in}}=1$, dashed] and
squeezed by 3 dB [$S_{X^{in}}=0.5$, plain], under the same pumping
conditions ($C=100$, $\Gamma_E=10\gamma$, $\gamma_0=\gamma/1000$).
The transfer efficiency is $\eta=0.91$ in the second
case.}\label{fig2}
\end{figure}

From Eqs. (\ref{Xzt}-\ref{jxzt}) it is possible to compute the
variances of the ground state spin coherence. The general method to
perform these calculations is detailed in the Appendix. It yields
the spectrum of the collective spin-squeezed coherence $J_x$ of the
squeezed component, $S_{J_x}(\omega)$, defined as \beqr\nn \langle
J_x(\omega)J_x(\omega')\rangle=2\pi\delta(\omega+\omega')S_{J_x}(\omega).\eeqr
The atomic spectrum is found to be the sum of three contributions:
\beqr\label{spectreJx}
S_{J_x}(\omega)&=&\frac{N}{4}\left[B_{f}(\omega)S_{X^{in}}+B_{coh}(\omega)+B_{spin}(\omega)\right]\eeqr
The first term in (\ref{spectreJx} is the coupling with the incident
squeezed vacuum fluctuations and quantifies how much of the incident
field squeezing is transferred to the spin, whereas $B_{coh}$ and
$B_{spin}$ give the contribution of, respectively, the atomic noise
resulting from spontaneous emission, and the atomic noise due to the
loss of coherence in the ground state. Integrating over frequency
yields the sought variance \beqr\nn \Delta J^2_x&\equiv
&\int\frac{d\omega}{2\pi}S_{J_x}(\omega),\eeqr the exact expression
of which is not reproduced here. However, when the incident field is
a (coherent) vacuum state - $S_{X^{in}}=1$ - the atoms are in a
coherent spin state. This implies a simple relation between the
integrals of the $B$'s:
\beqr\nn\int\frac{d\omega}{2\pi}\left[B_f(\omega)+B_{coh}(\omega)+B_{spin}(\omega)\right]=1.\eeqr
In the case of an amplitude squeezed input one can then measure the
efficiency of the squeezing transfer by comparing the atomic
squeezing to that of the incident field: \beqr\nn
\eta\equiv\frac{1-(\Delta J^2_x)/(N/4)}{1-S_{X^{in}}}.\eeqr $\eta=1$
thus corresponds to perfect transfer, $\eta=0$ to no transfer at
all. Using the previous relations the efficiency is equal to
\beqr\label{etaeitex} \eta
=\int\frac{d\omega}{2\pi}B_f(\omega)=\int\frac{d\omega}{2\pi}\frac{C\Gamma_E\gamma^2}
{|D|^2}\frac{|1-e^{-\alpha}|^2}{|\alpha|^2}\eeqr with
$D=(\gamma_0-i\omega)(\gamma-i\omega)+\Omega^2$. For most relevant
situations, however, the field and the optical dipole evolve rapidly
compared to the ground state coherence, so that it is possible to
adiabatically eliminate them in (\ref{Xzt}-\ref{jxzt}) and retrieve
simple analytical expressions for the atomic spectrum and variance.
In Fig. \ref{fig2} we represent a typical atomic noise spectrum for
typical experimental parameters. The atomic spectrum has a width
proportional to $\Gamma_E/\sqrt{C}$ for large $C$. In order to
maximize the transfer efficiency the pumping must be chosen in the
regime $\gamma_0\ll\Gamma_E/\sqrt{C}\ll\gamma$. The efficiency can
then be shown to be \beqr\label{etaeit} \eta\simeq
1-\frac{\sqrt{2/\pi}}{\sqrt{C}}-\frac{C\gamma_0}{\Gamma_E}
\hspace{1cm}(C\gg1,\gamma_0\ll\gamma)\eeqr A very good efficiency
can thus be reached for a large cooperative behavior, and, as in the
cavity scheme, the cooperativity is again the relevant parameter to
quantify the transfer efficiency. Note also that the ground state
decay rate can also contribute to degrade the squeezing when the
number of atoms grows large.

\subsection{Raman interaction}

We now consider a situation in which both fields are strongly
detuned with respect to the one-photon resonance
($\Delta_i\gg\gamma$), but the two-photon resonance is maintained
($\delta=0$), using a small longitudinal magnetic field for
instance. In this Raman interaction one can eliminate the optical
dipole and write simplified equations for the ground state coherence
and the field \beqr\nn \left(\dt +c\dz\right) X(z,t) &=&
-2\tilde{g}N j_y(z,t)\\\nn \dt j_y(z,t) &=&
-(\gamma_0+\Gamma_R)j_y+\frac{\tilde{g}}{2}X+f_{j_y},\eeqr where
$\Gamma_R=\gamma\Omega^2/\Delta^2$ is the Raman optical pumping rate
(assumed much smaller then $\gamma$) and $\tilde{g}=g\Omega/\Delta$
is the effective atom-field coupling constant. Note that in this
case the amplitude fluctuations are coupled to those of $j_y$.
Following a method analogous to the previous section the outgoing
field noise spectrum can be written as \beqr\label{spram}
S_{X^{out}}(\omega)=1-\left[1-S_{X^{in}}\right]e^{-(\alpha'+\alpha'^*)},\eeqr
with \beqr\nn \alpha'(\omega)=-i\frac{\omega
L}{c}+\frac{C\Gamma_R}{\Gamma_R+\gamma_0-i\omega}.\eeqr
\begin{figure}[h]
\includegraphics[width=8cm]{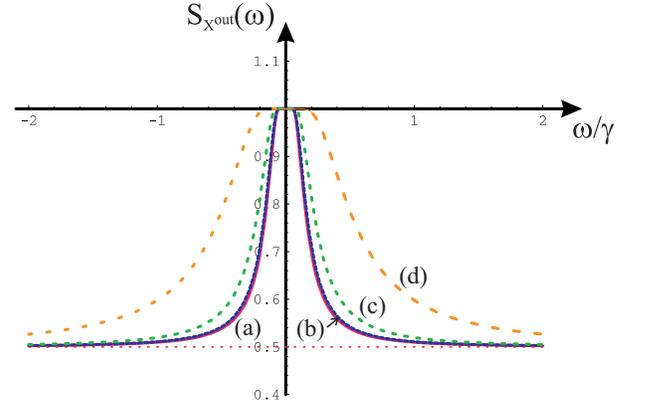}
\caption{Outgoing field noise spectrum in a Raman situation for the
same values of $\gamma_0$ as in Fig. \ref{fig1}. Parameters:
$\Gamma_R=\gamma/100$, $C=100$.} \label{fig3}
\end{figure}
The spectrum, plotted in Fig. \ref{fig3}, is radically different
from the EIT one; the squeezing is now absorbed around
zero-frequency on a width \beqr\nn\Delta\omega'\simeq
\sqrt{\frac{2}{\ln 2}}\sqrt{C\Gamma_R(\Gamma_R+\gamma_0)},\eeqr and
the spectrum broadens when the number of atoms (or the propagation
length) is increased. For higher frequencies the field comes out
unchanged. Note also that, the optical pumping rate being kept
constant, the width of the spectrum noticeably depends on the value
of the ground state decay rate: as can be seen from Fig. \ref{fig3}
the central absorption peak width increases with $\gamma_0$ as soon
as $\gamma_0\sim\Gamma_R$.

\begin{figure}[h]
\includegraphics[width=8cm]{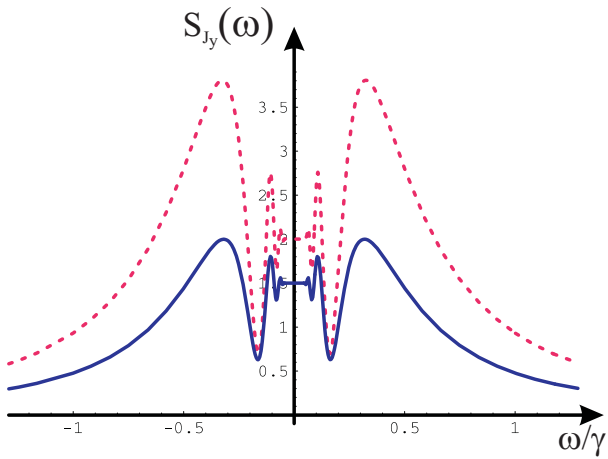}
\caption{Noise spectrum of the $y$-component of the spin for a Raman
interaction is in a coherent state [$S_{X^{in}}=1$, dashed] and
squeezed by 3 dB [$S_{X^{in}}=0.5$, plain]. Parameters: $C=100$,
$\Gamma_R=\gamma/100$, $\gamma_0=\gamma/1000$. The transfer
efficiency is $\eta'\simeq 0.91$ for a 3 dB-squeezed input, as in
EIT.} \label{fig4}
\end{figure}

Concerning the atoms, one finds an atomic spectrum for the spin
component coupled to $X^{in}$ as represented in Fig. \ref{fig4}.
Although it is rather different from the EIT spectrum the transfer
efficiency is remarkably similar \beqr\label{etaramex}
\eta'=\int\frac{d\omega}{2\pi}\frac{C\Gamma_R}{(\Gamma_R+\gamma_0)^2+\omega^2}\frac{|1-e^{-\alpha'}|^2}{|\alpha'|^2}.\eeqr
In this case the atomic noise spectrum width depends on
$\sqrt{C}\Gamma_R$, so that the good regime for quantum state
transfer is this time $\gamma_0\ll\sqrt{C}\Gamma_R\ll\gamma$. The
efficiency, as in EIT, increases to 100\% with the cooperativity as
$C^{-1/2}$, but shows a different sensitivity to ground state
decoherence: \beqr \label{etaram}\eta'\simeq
1-\frac{\sqrt{2/\pi}}{\sqrt{C}}\sqrt{1+\gamma_0/\Gamma_R}\eeqr for
$C\gg 1$ and $\gamma_0\ll\gamma$.

\section{Single pass \textit{vs} cavity}

It was shown in \ct{dantan1} that, if the atomic cloud was placed
inside a single-ended optical cavity, with an output coupling mirror
transmission $T$, a quasi-ideal mapping of the incident field is
possible either in EIT or Raman. If we first consider the EIT
situation, the atomic spectrum is Lorentzian-shaped with width
$\tilde{\gamma}_0=\gamma_0+\Gamma_E/(1+2C)$, whereas the field
exiting the cavity is squeezed by approximately the same amount as
the incident field for all frequencies. This is a strong difference
with the single pass scheme in which the outgoing field is squeezed
only in the transparency window, i.e. for low frequencies. This is
clearer when looking at the intracavity field fluctuations, $X$, and
relating them to those of the output field,
$X^{out}=\sqrt{T}X-X^{in}$: \beqr \nn
X(\omega)=\frac{2}{\sqrt{T}}\frac{1}{1+2C}\left[1+\frac{2C\tilde{\gamma}_0}{\tilde{\gamma}_0-i\omega}\right]
X^{in}+\frac{i\omega}{\tilde{\gamma}_0-i\omega}F,\eeqr with $F$ some
atomic noise operator. For frequencies $\omega\ll\tilde{\gamma}_0$,
$X\sim 2/\sqrt{T} X^{in}$, and the output field fluctuations are
those of the incident field: $ X^{out}\sim X^{in}$. This means that,
the medium being transparent in this frequency window, the
intracavity field is simply the incident field, and since there is
no field radiated by the atoms, the output field is the same as the
input. However, at high frequencies, the intracavity field
fluctuations are in $\mathcal{O}(1/C)$, and the output field is
equal to the reflected field: $X^{out}\sim - X^{in}$. Indeed,
outside the transparency window the incident field fluctuations are
absorbed by the atoms which radiate a field interacting
destructively with the incident field: $ X_r\sim -X^{in}$, so that $
X\varpropto X_r+X^{in}\simeq 0$. In contrast, in the single pass
scheme, this reflected field contribution to the output field is of
course not present, so that the squeezing disappears outside the
transparency window.\\

If we now compare with the Raman situation, this frequency
dependence is opposite. The intracavity field fluctuations can be
written as \beqr \nn
X=\frac{2}{\sqrt{T}}\frac{\Gamma_R-i\omega}{\tilde{\gamma}_0-i\omega}
X^{in}+\frac{F'}{\tilde{\gamma}_0-i\omega},\eeqr where the effective
atomic decay rate is now $\tilde{\gamma}_0=\gamma_0+(1+2C)\Gamma_R$
and, again, $F'$ some atomic noise operator. At low frequencies, one
has $ X\simeq 0$, so that the output fluctuations are those of the
reflected field, $ X^{out}\sim - X^{in}$. On the contrary, for
frequencies $\omega\gg\tilde{\gamma}_0$, $ X\sim 2/\sqrt{T}X^{in}$
and $ X^{out}\sim X^{in}$. This is again in good agreement with what
was found for the single pass scheme.\\

Coming now to the atoms, a noticeable difference is the transfer
efficiency: in the cavity scheme the efficiency increases to 1 as
$1/C$, whereas, in the single pass scheme, the increase is slower -
in $1/\sqrt{C}$. Physically, it means that the atom-field
interaction in a cavity with $N$ atoms and an output mirror
transmission $T$ is not equivalent to a single pass interaction with
$N/T$ atoms, even though the cooperativities are then equal in both
cases [$C=g^2N/T\gamma$ in a cavity \ct{dantan1}]. This is naturally
due to the fact that the incident squeezing is recycled inside the
cavity on each round trip, whereas, in the single pass scheme, the
atoms "see" less and less squeezing along the propagation pass. This
accounts for the fact that the cavity scheme intrinsic atomic noise
decreases as $1/N$ and as $1/\sqrt{N}$ in a single pass scheme.
First, in an EIT cavity configuration and for $C\gg 1$,
$\Gamma_E\gg\gamma_0$, the efficiency can be written as \ct{dantan1}
\beqr\nn \eta &=&
\frac{2C}{1+2C}\frac{\Gamma_E/(1+2C)}{\gamma_0+\Gamma_E/(1+2C)}\\\label{etacav}\simeq
&&1-\frac{1}{1+2C}-\frac{(1+2C)\gamma_0}{\Gamma_E}\eeqr Comparing
Eq. (\ref{etacav}) with Eq. (\ref{etaeit}) it is clear that the
difference in efficiency comes from the sensitivity to noise coming
from spontaneous emission, damped by a factor $(1+2C)$ in a cavity
configuration and by $\sqrt{C}$ in single pass. Note, however, that
the robustness of the mapping operation with respect to ground state
decoherence is the same in cavity and single pass, because the
absorption is then linear in the number of atoms effectively seen by
the field, i.e. proportional to $C$ in both cases. This drawback can
be overcome with the use of a buffer gas, which can significantly
reduce the ground state decay rate and increase the transfer
efficiency \ct{hau}.

For a Raman interaction this sensitivity is less crucial, since its
effect is to reduce the cooperativity from $C$ to
$C\Gamma_R/(\Gamma_R+\gamma_0)$, as can be seen from (\ref{etaram}).
This difference with EIT comes from the fact that the range of
frequencies involved in the Raman interaction is broader [see Fig.
\ref{fig4}] than in EIT. In the latter the atomic noise reduction is
greater around zero frequency where the effect of ground state
decoherence is the most important. In Fig. \ref{fig4} is plotted the
transfer efficiency in both schemes, when the cooperativity is
varied. It is worth noticing that, even though the increase in
efficiency is slower in single pass than in cavity, an excellent
mapping - $\eta \sim 100\%$ - can be achieved for all these schemes
when the cooperativity is high enough.

\begin{figure}[h]
\includegraphics[width=8cm]{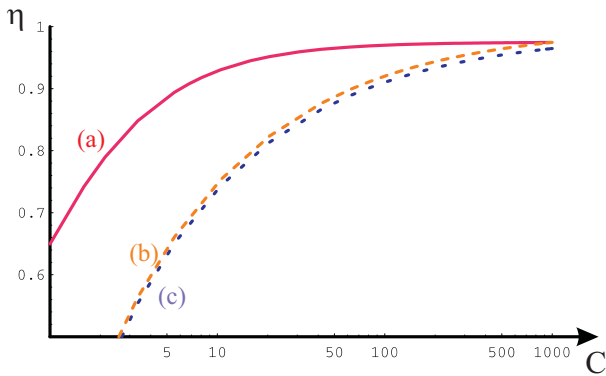}
\caption{Mapping efficiency versus cooperativity in a cavity scheme
(a) and a single pass interaction [(b) Raman, (c) EIT]. Parameters:
$\gamma_0=\gamma/1000$. }\label{fig5}
\end{figure}

\section{Discussion and conclusion}

It is therefore possible to map almost perfectly a squeezed vacuum
field state onto a ground state spin coherence, results already
predicted for a cavity scheme in \ct{dantan1}. The same decrease in
transfer efficiency is found in EIT or Raman as compared to the
cavity schemes under the same conditions. However, good quality
transfer remains possible for very realistic parameters. Most
properties relative to the transfer processes stress the importance
of the cooperative behavior of the atoms. If qualitatively the
conclusions drawn in the cavity scheme remain valid in a single pass
approach, quantitatively, however, the difference of scaling with
the cooperativity shows that the cavity scheme is more efficient in
many ways: writing and readout time, mapping efficiency, robustness
with respect to spontaneous emission, etc. It is also interesting to
note the differences and similarities between Raman and EIT
interactions. Good tests of this theory could be provided by
outgoing field noise measurements, such as those performed in
\ct{akamatsu}. Limitations may arise from the imperfections of the
one-dimensional theory. For instance, diffraction effects or the
issue of matching between the field and atomic modes are expected to
play an important role \ct{duan2,polzik2} when the Fresnel number is
not unity or when the plane wave approximation for the field is no
longer valid. In \ct{dantan2} it was shown that this quantum state
exchange mechanism also extends to EPR-entangled fields interacting
with two ensembles in cavities. The calculations of this paper
naturally extend to such states in single pass interactions,
provided suitable interaction parameters are chosen. Most ideas
developed in \ct{dantan1,dantan2,dantan3} are also transposable to
single pass interaction, which should simplify the manipulation and
storage of quantum states in atom-field quantum communication
networks.

\begin{acknowledgments}
In the course of redaction of this paper we became aware of similar
work on EIT by A. Peng \textit{et al.} \ct{peng}, who reach the same
conclusions about the outgoing field spectrum in EIT.
\end{acknowledgments}

\appendix
\section{Atomic spectrum calculation}
From Eqs. (\ref{Xzt}-\ref{jxzt}) it is possible to compute the
variances of the ground state coherence. We would like to stress the
method to solve these space- and time-dependent coupled differential
equations, method which is actually quite general and may be applied
to other situations. The idea is to perform a Fourier transform in
time and a Laplace transform in space, in order to have a simple
linear system. We standardly define the Laplace transform of $f(z)$
as \beqr\nn f[s]=\int_{0}^{\infty}e^{-sz}f(z)dz,\eeqr and the
Fourier transform of $g(t)$ as \beqr\nn
g(\omega)=\int_{-\infty}^{\infty}e^{i\omega t}g(t)d\omega.\eeqr The
system (\ref{Xzt}-\ref{jxzt}) then becomes \beqr\nn
(-i\omega+cs)X[s,\omega]&=&c X(0,\omega)-2gN\sigma_y[s,\omega]\\\nn
(\gamma-i\omega)\sigma_y[s,\omega]&=&\Omega
j_x[s,\omega]+\frac{g}{2} X[s,\omega]+f_{\sigma_y}[s,\omega]\\\nn
(\gamma_0-i\omega) j_x[s,\omega]&=&-\Omega
\sigma_y[s,\omega]+f_{j_x}[s,\omega]\eeqr From these equations one
deduces $j_x[s,\omega]$: \beqr \nn j_x[s,\omega]&=&\frac{B_1}{s+s_0}
X^{in}(\omega)\\\nn&&+B_2\frac{s-b_2}{s+s_0}f_{\sigma_y}[s,\omega]
+B_3\frac{s-b_3}{s+s_0}f_{j_x}[s,\omega]\eeqr with
$X^{in}(\omega)=X(0,\omega)$, $B_1= -\frac{g\Omega/2}{D}$,
$B_2=\frac{-\Omega}{D}$, $B_3=\frac{\gamma-i\omega}{D}$,
$b_2=i\omega/c$, $b_3=i\omega/c-g^2N/(\gamma-i\omega)$,
$D=(\gamma_0-i\omega)(\gamma-i\omega)+\Omega^2$, and $s_0=
-i\frac{\omega}{c}D+g^2N(\gamma_0-i\omega)$. Using inverse Laplace
transforms one then gets the fluctuations of the atomic operators at
position $z$ \begin{align}&\nn
j_x(z,\omega)=B_1e^{-s_0z}X^{in}(\omega)\\\nn
&+B_2\left[f_{\sigma_y}(z,\omega)-(s_0+b_2)\int_0^zdz'e^{-s_0(z-z')}f_{\sigma_y}(z',\omega)\right]\\\nn
&+B_3\left[f_{j_x}(z,\omega)-(s_0+b_3)\int_0^zdz'e^{-s_0(z-z')}f_{j_x}(z',\omega)\right]\end{align}
Finally, integrating over $z$ yields the collective spin
fluctuations \begin{align}
&J_x(\omega)=B_1N\frac{1-e^{-\alpha}}{\alpha}X^{in}(\omega)\nn\\
&\nn+NB_2\int_0^L\frac{dz}{L}\left[f_{\sigma_y}(z,\omega)-\lambda_2\int_0^zdz'e^{-s_0(z-z')}f_{\sigma_y}(z',\omega)\right]\\\nn
&+NB_3\int_0^L\frac{dz}{L}\left[f_{j_x}(z,\omega)-\lambda_3\int_0^zdz'e^{-s_0(z-z')}f_{j_x}(z',\omega)\right]\end{align}
with $ \lambda_i=s_0+b_i$ ($i=2,3$). Using the correlation functions
of the $f$'s and of the incident field one deduces the expressions
of the functions $B_f$, $B_{coh}$, and $B_{spin}$ of the atomic
field spectrum (\ref{spectreJx}).


\bigskip

\begin{thebibliography}{99}

\bibitem{lukin} M.D. Lukin, Rev. Mod. Phys. \textbf{75}, 457
(2003).

\bibitem{duan} L.M. Duan, J.I. Cirac, P. Zoller, and E.S. Polzik, Phys. Rev.
Lett. \textbf{85}, 5643 (2000); L.M. Duan, M.D. Lukin, J.I. Cirac,
and P. Zoller, Nature \textbf{414}, 413 (2001).

\bibitem{harris} S.E. Harris, Phys. Today \textbf{50}, 36 (1997).

\bibitem {hau} M.M. Kash \textit{et al.}, Phys. Rev. Lett.
\textbf{82}, 5229 (1999); C. Liu, Z. Dutton, C.H. Behroozi, and L.V.
Hau, Nature \textbf{409}, 490 (2001); D.F. Phillips, A.
Fleischhauer, A. Mair, and R.L. Walsworth, and M.D. Lukin, Phys.
Rev. Lett. \textbf{86}, 783 (2001); M. Bajcsy, A.S. Zibrov, and M.D.
Lukin, Nature \textbf{426}, 633 (2004).

\bibitem{fleischhauer} M.D. Lukin and M. Fleischhauer, Phys. Rev. Lett. \textbf{84},
5094 (2000); M. Fleischhauer and M.D. Lukin, Phys. Rev. A
\textbf{65}, 022314 (2002).

\bibitem {julsgaard} B. Julsgaard, A. Kozhekin, and E.S. Polzik, Nature \textbf{413}, 400
(2001); C. Schori, B. Julsgaard, J.L S\o rensen, and E.S. Polzik,
Phys. Rev. Lett. \textbf{89}, 057903 (2002).

\bibitem{julsgaard2} B. Julsgaard \textit{et al.}, quant-ph/0410072.

\bibitem{kuzmich} A. Kuzmich, K. M\o lmer, and E.S. Polzik, Phys. Rev. Lett.
\textbf{79}, 4782 (1997); A.E. Kozhekin, K. M\o lmer, and E.S.
Polzik, Phys. Rev. A \textbf{62}, 033809 (2000).

\bibitem {dantan1} A. Dantan and M. Pinard,
Phys. Rev. A \textbf{69}, 043810 (2004).

\bibitem{dantan2} A. Dantan, A. Bramati, and M. Pinard, Europhys. Lett.
\textbf{67}, 881 (2004).

\bibitem{raymer} M.G. Raymer and J. Mostowski, Phys. Rev. A
\textbf{24}, 1980 (1981).

\bibitem{duan2} L.M. Duan, J.I. Cirac, and P. Zoller, Phys. Rev. A \textbf{66}, 023818 (2002).

\bibitem{fleischhauer95} M. Fleischhauer and T. Richter, Phys. Rev.
A \textbf{51}, 2430 (1995); I.H. Deutsch and J.C. Garrison, Phys.
Rev. A \textbf{43}, 2498 (1991); A. Siberfarb and I.H. Deutsch,
Phys. Rev. A \textbf{68}, 013817 (2003).

\bibitem{dantan3} A. Dantan, N. Treps, A. Bramati, and M. Pinard,
quant-ph/0407258.

\bibitem{nusszenveig} C.L. Garrido-Alzar, L.S. Cruz, J.G.
Aguirre-G\'{o}mez, M. França Santos, and P. Nusszenveig, Europhys.
Lett. \textbf{61}, 485 (2003).

\bibitem{akamatsu} D. Akamatsu, K. Akiba, and M. Kozuma, Phys. Rev.
Lett. \textbf{92}, 203602 (2004).

\bibitem{ueda}  M. Kitawaga and M. Ueda, Phys. Rev. A \textbf{47},
5138 (1993); D.J. Wineland, J.J. Bollinger, W.M. Itano, and D.J.
Heinzen, Phys. Rev. A \textbf{50}, 67 (1994).

\bibitem{molmer} U.V. Poulsen and K. M\o lmer, Phys. Rev. Lett.
\textbf{87}, 123601 (2001).

\bibitem{polzik2} J.H. M\"{u}ller \textit{et al.}, quant-ph/0403138.

\bibitem{peng} A. Peng, M. Johnsson, and J.J. Hope,
quant-ph/0409183.

\end{thebibliography}
\end{document}